\documentstyle[aps,preprint]{revtex}
\input psfig
\begin{document}
\draft
\preprint{Fermilab-Pub-94/295-A}
\date{August 30, 1994}
\title{Kinetics of Bose-Condensation}

\author{D. V.\ Semikoz and I. I.\ Tkachev}
\address{NASA/Fermilab Astrophysics Center
Fermi National Accelerator Laboratory, Batavia, IL~~60510}
\address{and Institute for Nuclear Research
Russian Academy of Sciences, Moscow 117312, Russia}

\maketitle

\begin{abstract}
The process of condensation in the system of scalar Bosons with weak
$\lambda \phi^4$ interaction is considered. Boltzmann kinetic  
equation is
solved numerically. Bose condensation proceeds in two stages:
At the first stage condensate is still absent but there is non-zero   
inflow
of particles towards $\vec{{\bf p}} = 0$ and the distribution  
function
at $\vec{{\bf p}} = 0$ grows from finite values to infinity.
At the second stage there are two components, condensate and  
particles,
reaching their equilibrium values. We show that the evolution in both  
stages
proceeds in a self - similar way and find the time needed for  
condensation,
which is finite.
\end{abstract}

\narrowtext

It is a fundamental result of quantum statistics of Bosons that above  
a certain
critical density all added particles must enter the ground state:  
Bose-Einstein
condensate forms. The kinetics of this process is a very interesting  
problem.
One can reach the Bose condensation gradually decreasing temperature  
in a
sequence of equilibrium states. An appropriate description of this is  
given by
the well-known kinetics of second order phase transitions. On the  
other hand,
when the conditions for the formation of Bose condensate appear, the  
system
can be far from the equilibrium. Recently this problem attracted  
particular
interest in connection with exciting prospects for the experimental  
observation
of Bose condensation in a very cold atomic samples, e.g. in a gas of
spin-polarized atomic hydrogen \cite{bprl} or in alkali-metal vapors
\cite{m90}. Another interesting application of Bose kinetics is  
rather far from
the laboratory experiments and is related to the problem of Bose  
stars
formation \cite{it91,kt93} from the dark matter in the universe.

The question of the time evolution of weakly interacting Bose gas was  
addressed
in a number of papers. In earlier treatments an ideal Bose gas was  
coupled to a
thermal bath with infinite heat capacity \cite{ih76,ly77}. Small  
energy
exchange was assumed and after several other approximations a  
Fokker-Planck
type equation was obtained. Levich and Yahkot \cite{ly77} calculated
analytically that the time for condensation is infinite in this  
situation. By
including Boson-Boson interactions they later \cite{ly78} found  
solution which
describes explosive appearance of a condensate, but they concluded  
that this
effect could have been an artifact of their approximations.

Snoke and Wolfe in Ref. \cite{sw89} undertook direct numerical  
integration of
Boltzmann kinetic equation.  Although this calculation demonstrated  
the
restructuring of the distribution function, the appearance of a Bose  
condensate
was not detected. Their approach is close to ours but in comparison  
with the
Ref. \cite{sw89} we perform numerical integration in much wider  
dynamical range
of relative energies and densities and we directly analyse the  
behavior the
distribution function.

Another result recently reported in Ref. \cite{s91} states that the  
time
required for condensation is $\sim T_c$, where $T_c$ is the  
temperature of Bose
condensation. This result seems to be incorrect (or at least it can  
not be
applied to all temporal stages of gas evolution) since it is  
insensitive to the
interactions. Clearly, in the limit of zero couplings the  
distribution of
particles does not evolve and the relaxation time has to be infinite.

Analytical study of Bose condensation was performed recently in the  
paper
\cite{kss92}. Three different regimes of evolution where identified.  
It was
argued that in the kinetic region in non-linear regime the  
distribution
function has to be a power law $f \propto \varepsilon ^{-7/6}$. This  
power low
is well known in the theory of plasma turbulence \cite{k41,z66}. We  
indeed
observe the tendency to this law in our numerical simulations, but  
the system
never reaches it. Moreover, this distribution became destroyed with  
condensate
appearance, contrary to the assumptions of Ref.\cite{kss92}. What  
concerns
condensation time, only rough dimensional estimations were done in  
Ref.
\cite{kss92}.

We found that Bose condensation process can be divided on two stages.  
During
the first part of the first stage the self-similar solution forms and  
then the
distribution function reaches infinity at $\vec{p}=0$ in a  
self-similar way. We
can not observe in principle the actual build up of the coherence in  
the
frameworks of Boltzmann equation. But we have to conclude that after  
the
distribution function became infinite at zero momentum, the  
condensate had
formed. While due to fluctuations the width of the coherent region  
can be
finite in momentum space, we simply model the coherent field by  
$N_c\,\delta
(\vec{{\bf p}})$  with initially infinitesimally small but then  
growing
amplitude $N_c(t)$. We found that in this second stage the evolution  
also
proceeds in a self-similar fashion and found the durations of both  
stages.

While our prime interest and motivations for this work was connected  
to the
physics of Bose star, the problem of condensate formation with  
account for the
gravitational field was never approached and we do not attempt it  
here.
Instead, we are solving kinetic equations in flat space-time and only  
range of
parameters of the initial distribution reflect the virial equilibrium  
of
self-gravitating system.

We consider the system of scalar bosons with 4-particle  
self-interaction.
The field-theoretical Lagrangian which we bear in mind is
$L = (\partial_{\mu}\phi)^2/2 - m^2\phi^2/2 -\lambda \phi^4/4!$.
Time development of the quantum state which contain well defined and
large number of particles can be
 adequately described by the Boltzmann kinetic equation which governs  
the
evolution of one-particle distribution function, $f(\vec{{\bf p}})$:
\begin{equation}
\frac{df (\vec{{\bf p}}_1)}{dt} = \frac{\pi^4}{m^4} \int
|M_{fi}|^2 F(f)\delta^4(\sum_i p_{\mu i})  \prod_{i=2}^4
\frac{d^3\vec{{\bf p}}_i}{(2\pi)^3 }  ~,
\label{kin}
\end{equation}
where
\begin{equation}
F(f)=[1+f_1]\,[1+f_2]f'_1f'_2-[1+f'_1]\,[1+f'_2]f_1f_2~,
\label{Fun}
\end{equation}
and $f_i \equiv f(\vec{p}_i)$, $f'_i \equiv f(\vec{p}_i\, ')$.
The equilibrium solution of the kinetic equation
is the Bose-Einstein distribution function
\begin{equation}
f(\vec{{\bf p}})=\frac{1}{\exp{((\varepsilon-\mu)/T)}-1} + (2\pi  
)^3N_c \delta
(\vec{{\bf p}}) ~,
\label{stat}
\end{equation}
where $\varepsilon$ is the particle energy, $\mu$ is the chemical  
potential,
$T$ is the
temperature of the final state   and $N_c$ is the
number density of particles in condensate.

In  what follows we  shall consider isotropic initial distribution
$f=f(\varepsilon)$.
In our case  the matrix element is given by
$|M_{fi}|=  \lambda^2$
and the kinetic equation for the case without condensate, $N_c = 0$,
can be rewritten in the form

\begin{equation}
\frac{df (\varepsilon_1)}{dt} = \frac{\lambda^2}{64\pi^3m} \int \int
  F(f) \frac{D}{p_1} d\varepsilon'_1 d\varepsilon'_2 \equiv I_P~,
\label{kin2}
\end{equation}
where $D \equiv \text{min} [p_1,p_2,p'_1,p'_2]$ and $\varepsilon_2 =
\varepsilon'_1+\varepsilon'_2-\varepsilon_1 $ in arguments of $F(f)$.
The integration should be done over the region  
$\varepsilon_1<\varepsilon'_1
<\infty$, $\varepsilon_1-\varepsilon'_1<\varepsilon'_2 <\infty$.

After the moment of condensate  formation the kinetic equation  
(\ref{kin2})
is inappropriate for numerical integration anymore,  the finite  
number of
particles in condensate corresponds to the infinite value of the  
distribution
function at zero energy.
In order to describe the system of particles interacting with the  
condensate
we divide the distribution function into two pieces:
$\tilde{f} = f(\varepsilon,t) + (2\pi )^3N_c(t) \delta^3 (\vec{{\bf  
p}})$,
where the first term corresponds to the "gas" of particles and the
second one describe the condensate.
Substituting this function into the original kinetic equation  
(\ref{kin})
we obtain
\newpage
\begin{mathletters}
\label{kin3}
\begin{equation}
\dot{N_c}(t) = \frac{\lambda^2N_c(t)}{64\pi^3m} \int_0^\infty  
d\varepsilon'_1
d\varepsilon'_2 [f'_1f'_2-f^{ }_2(f'_1+f'_2)]~,
\label{equationa4}
\end{equation}
\begin{equation}
\dot{f}(\varepsilon_1) = I_P +\frac{N_c(t)\lambda^2}{32\pi m^2p_1}  
\left(
\int^{\varepsilon_1}_0
[(f'_2-f^{ }_1)f'_1-f'_2f^{ }_1]d\varepsilon'_2 \right.
 + \left. 2 \int_{\varepsilon_1}^\infty [(f'_2-f^{ }_1)f^{  
}_2+f'_2f^{ }_1]
d\varepsilon'_2  \right) ~.
\label{fdot4}
\end{equation}
\end{mathletters}

In general, after the condensate formation (and  at large particle  
densities
even before) the description in terms of quasiparticles rather then  
particles
is more appropriate. For example, the kinetic equation,  
Eq.(\ref{kin3}),
does not include the processes where one of the incoming and one of  
the
outcoming particles has zero momentum, $p'_2 = p_2 = 0$. This process  
does not
contribute to the collision integral directly, i.e. it does not  
change
the distribution of particles over energies, but it does change the
effective particle mass.

However, in many cases those effects are insignificant and we still  
can work in
terms of particles. The quantitative
arguments are the following. The effective mass of quasiparticles in  
the
presence of condensate (or in dense medium) is $m_{\text{eff}} = m^2
+ \lambda\phi_c^2/6 = m^2 + \lambda n/m$. We can still use the
description in  terms of particles if the second term in the sum is  
much
smaller than the first one. Since $n \sim m^3 (\Delta \upsilon)^3f_0$
we obtain $\lambda f_0(\Delta \upsilon)^3 \ll 1$, where $\Delta  
\upsilon$ is
characteristic velocity dispersion.  In the case of axion  
miniclusters, for
example,  we have \cite{kt93} $\lambda f_0\sim 10^{-5}$; $\Delta  
\upsilon \sim
10^{-8}$, and the description in terms of particles is perfectly  
good.

As an initial distribution we choose the function $f(\varepsilon)$  
which has
the maximum at $\vec{{\bf p}}=0$. In general, such  distribution  
function can
be characterized by means of three major parameters:
(1) The overall amplitude $f_0$. In what follows we define
$f_0=f(\varepsilon=0)$.
(2) The energy scale $\varepsilon_0$ where the distribution function
became twice smaller, $f(\varepsilon_0)=f_0/2$.
(3) The effective width, $\Gamma$, of the region over which the  
distribution
function varies rapidly.
More specifically, we choose the initial distribution function to be  
of the
form:
\begin{equation}
f(\varepsilon)=\frac{2
f_0}{\pi}\arctan{\left[\exp{(\Gamma(1-\varepsilon/\varepsilon_0))}\right]} ~.
\label{fin}
\end{equation}
In what follows we shall measure the energy in units of  
$\varepsilon_0$ and the
distribution function in units of $f_0$,
i.e. initial distribution function will have the normalization
$f(\varepsilon=0)=1$.

We define the dimensionless time $\tau$ as \cite{it91}:
\begin{equation}
\tau \equiv \frac{\varepsilon_0^2 f_0 (1 + f_0) \lambda^2 }{64 \pi^3  
m }~ t~.
\label{time}
\end{equation}

The parameters  $\varepsilon_0$ and $f_0$ in the limit $f_0 \gg 1$  
after
rescaling will not enter the kinetic equation explicitly, but will  
define  the
time scale.
In terms of $\tau$ there remains the
weak dependence of the relaxation time upon the initial shape  
parameter
$\Gamma$. All data presented in this paper will correspond to one and  
the same
value of $\Gamma = 5$.
With the initial shape of the distribution function being given, the   
two
scaling parameters $f_0$ and $\varepsilon_0$  define also the  
parameters of
final equilibrium $\{ T,\mu, N_c \}$.
With $\Gamma =5$ the choice
$f_0 \leq f_{\text{crit}} \approx 2.8$ corresponds to the Bose-gas  
without
condensate, while
$f_0 > f_{\text{crit}}$ corresponds to condensate formation in the  
final
equilibrium state.

We shall consider here the  case  $f_0 > 2.8$ only. We shall simplify  
the
problem and consider $f_0 \gg 1$. In this case
we can  disregard $f^2$ terms in the function (\ref{Fun}),
which became
$F(f)=[f_1+f_2]f'_1f'_2-[f'_1+f'_2]f_1f_2$.

We integrated the kinetic equation in the energy interval
$10^{-9} < \varepsilon < 10$. We defined the distribution function on  
the
grid of 200 points equally spaced in the logarithm of energy and used  
the
 spline interpolation when calculating distribution function
at intermediate points. For each integration in collision integral
we had implied  Gauss algorithm. Particle and energy non-conservation  
was of
order $10^{-3}$ for the whole time of integration.

Results of numerical integration of the kinetic equation (\ref{kin2})
are presented in Fig. 1, where we plot the distribution function at  
different
moments of time. We have arranged the output each time when
$f(\varepsilon = \varepsilon_{\text{min}},t)$ had increased by one  
order of
magnitude. The most striking feature of this plot is self-similar  
character of
the evolution.
The distribution function has the "core" where $f(\varepsilon)  
\approx
\text{const}$ and the radius of "core"
decreases with time while the value of $f(\varepsilon)$ in the core  
grows.
Outside the core the distribution function is the power law  
$f(\varepsilon )
\propto \varepsilon^{-\alpha }$ and does not depend upon time to a  
very good
accuracy.
Self-similar solutions exhibiting this kind of behavior can be  
parametrized as
\begin{equation}
f(\varepsilon,\tau) = A^{-\alpha}(\tau) f_s(\varepsilon/A(\tau))~,
\label{fst}
\end{equation}
where it is assumed $A(\tau) \rightarrow 0$ with the increase of time
and we always can choose the normalization $f_s(0) = 1$.
It is possible to find the time dependence of distribution function  
at
$\varepsilon = 0$ at late times analytically, using self-similarity   
of the
solution.

Substituting parametrization (\ref{fst}) in the kinetic equation
we obtain:
\begin{equation}
f(0,\tau)=[2 C (\tau_c-\tau)(\alpha-1)]^{-\alpha/(2(\alpha-1))} ~,
\label{fzero}
\end{equation}
and $f(0,\tau)$ has the pole at the finite moment of time  
$\tau=\tau_c$,
$f(0,\tau_c)=\infty$.

The value of logarithmic derivative of $f(\varepsilon)$ is plotted in  
Fig. 1b.
The dotted line in Figs. 1 corresponds to the limiting value
$\alpha = 7/6$. This power law corresponds to the stationary solution  
of the
kinetic equation (\ref{kin2}), see Ref. \cite{z66}. But this value is  
never
reached prior to the moment of condensate formation (after that  
moment the
character of evolution completely changes). Rather, with the boundary
condition $df/d\varepsilon =0$ at $\varepsilon=0$ the power law on  
the tail is
$\alpha \approx 1.24 $ (with appropriate boundary conditions our code  
correctly
finds the root $\alpha =7/6$ ). And the function Eq. (\ref{fzero}) is  
a good
fit to our data with $\alpha \approx 1.24 $ and $\tau_c \approx 19$.

We conclude, that after the solution have
reached self-similar form, the time dependence of the distribution
function at zero momentum is given by
$f(0,\tau) \propto (\tau_c - \tau)^{-2.6}$,
which reaches infinity at finite time and means the onset of  
condensation.

Though the system does not reach the limiting value of $\alpha =7/6$,  
at least
this power law is the root of the equation $I_P = 0$  \cite{z66}.   
Because of
that we selected the function
$f(\varepsilon) \propto \varepsilon^{-7/6}$ at $\varepsilon < 1$
as an initial condition while integrating Eqs. (\ref{kin3}). We took  
initially
$N_c \ll N_{\text{tot}}$ (the particular choice for $N_c$, as far as  
this
condition is satisfied, is insignificant).

Results of numerical integration of the system (\ref{kin3}) are
presented in Fig. 2. The dashed line  corresponds to the initial
distribution. Solid lines correspond to the distribution
function  at different moments of time. Basically, evolution
proceeds in the following way. First, the power law
$f(\varepsilon) \propto 1/(\varepsilon)^{7/6}$ changes to the law
 $f(\varepsilon) \propto 1/\varepsilon$ at small energies.
And then this change propagates to the region of lager energies,
see Fig. 2. Later on the power law stays on the equilibrium value  
$\alpha = 1$,
but the amplitude of $f(\varepsilon)$ gradually decreases.

Again, we see that the essential parts of curves in Fig. 2 repeat  
itself
under translation from the left to the right and evolution is  
self-similar.
During this epoch (before "the wave of change" had reached  
exponential tail of
the initial distribution at $\varepsilon > 1$)
approximately $40\%$ of particles had condensed. And what is  
important,
the number of particles in condensate linearly grows with time at  
this
epoch, $N_c/N_{\rm tot}= B \tau$. This enables us to eliminate the  
ambiguity in
the initial value for $N_c$ since $B$ does not depend upon it.  
Indeed, in our
simulations which was done in a finite energy interval, during  
several first
iterations system self-adjusts: a proper profile of distribution  
forms while
condensate reaches particular value of $N_c$. We can disregard this  
period and
extrapolate curves in Fig. 2, $N_c(\tau )$ and self-similar character  
of the
evolution back in time and to the region of smaller energies.

We obtained $B \approx 2\times f_i(1)^2$ (initial normalization  
$f_i(1) = 1$
corresponds roughly to the magnitude of final distribution at  
$\varepsilon \sim
\varepsilon_{\rm min}$ in Fig. 1). For completeness we also had runs  
when we
used final distribution in Fig. 1 as initial condition for  
integration Eqs.
(\ref{kin3}). Condensation in this case decelerates and $N_c$ does  
not grow
lineraly, all is in accordance with excess magnitude of the  
distribution
function at small energies when $\alpha  \approx 1.24$.

We have studied numerically the kinetics of Bose condensation of the  
weakly
interacting Bose gas. The picture we observe differs even  
qualitatively from
the previous works. The distribution function of excess particles  
which
eventually has to form the condensate does not narrows in time  
gradually
approaching $\delta$- function, as it was found in Ref. \cite{ly77}  
assuming
small energy exchange per collision. Instead, power low profile $f  
\sim
\varepsilon^{-7/6}$ tend to form which corresponds to the constant  
flux of
particles in momentum space towards the condensate \cite{z66}. One  
could
expect, as it was done in Ref. \cite{kss92}, that this  very natural  
regime
will persist in the presence of condensate as well, till all excess  
particles
will inflow from the high energy tail and reside in the ground state.
Nevertheless, this is not the case also and at the moment the  
condensate
appears, it terminates this regime. Instead of this steady flow  
through the
whole energy interval, particles from all energy levels jump directly  
to the
condensate retaining during the major period of time the equilibrium  
shape of
the distribution function, $f \sim \varepsilon^{-1}$. The constant of
proportionality in this law gradually decreases till it reaches  
equilibrium
value.

We can not, in principle, observe the build up of coherence being in  
frameworks
of Boltzmann kinetic equation, but kinetic description has to be  
valid prior to
and after the very moment of condensate formation. We have seen that  
evolution
on both stages is self-similar. This allows us to obtain a number of  
useful
analytical relations, e.g. the time dependence of the distribution  
function
near the point of condensate formation, Eq. \ref{fzero}. We have  
found the
duration of both stages, which is finite and of order $\tau_c \sim  
20$.

We thank E. Kolb, V. Rubakov and G. Starkman for useful discussions.
This work was supported by DOE and NASA grant NAGW-2381 at Fermilab.
D.S. thanks the Astrophysics Department of FNAL for the hospitality  
during
this work.

\begin{figure}

\psfig{file=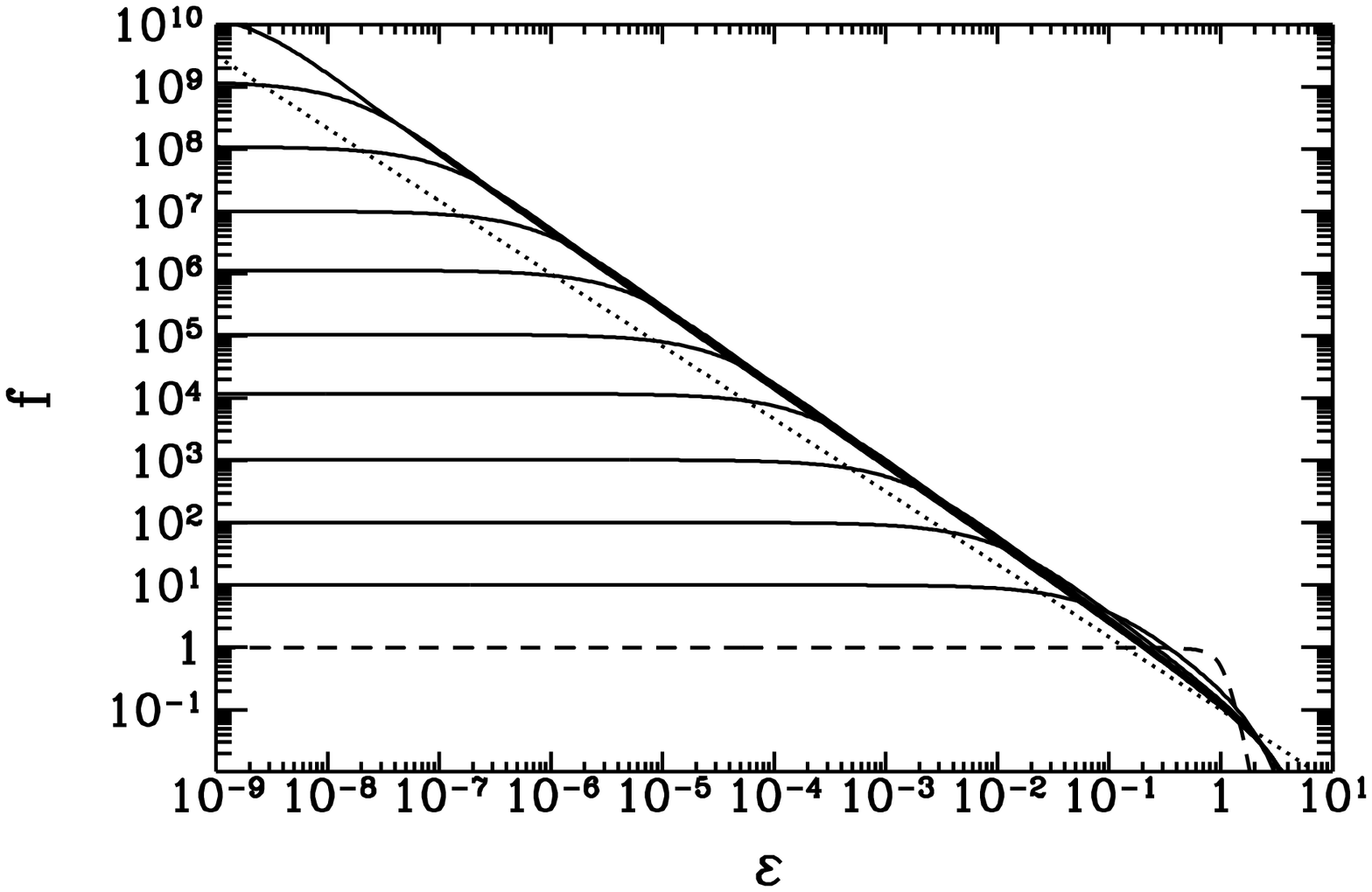,height=4in,width=6in}

\psfig{file=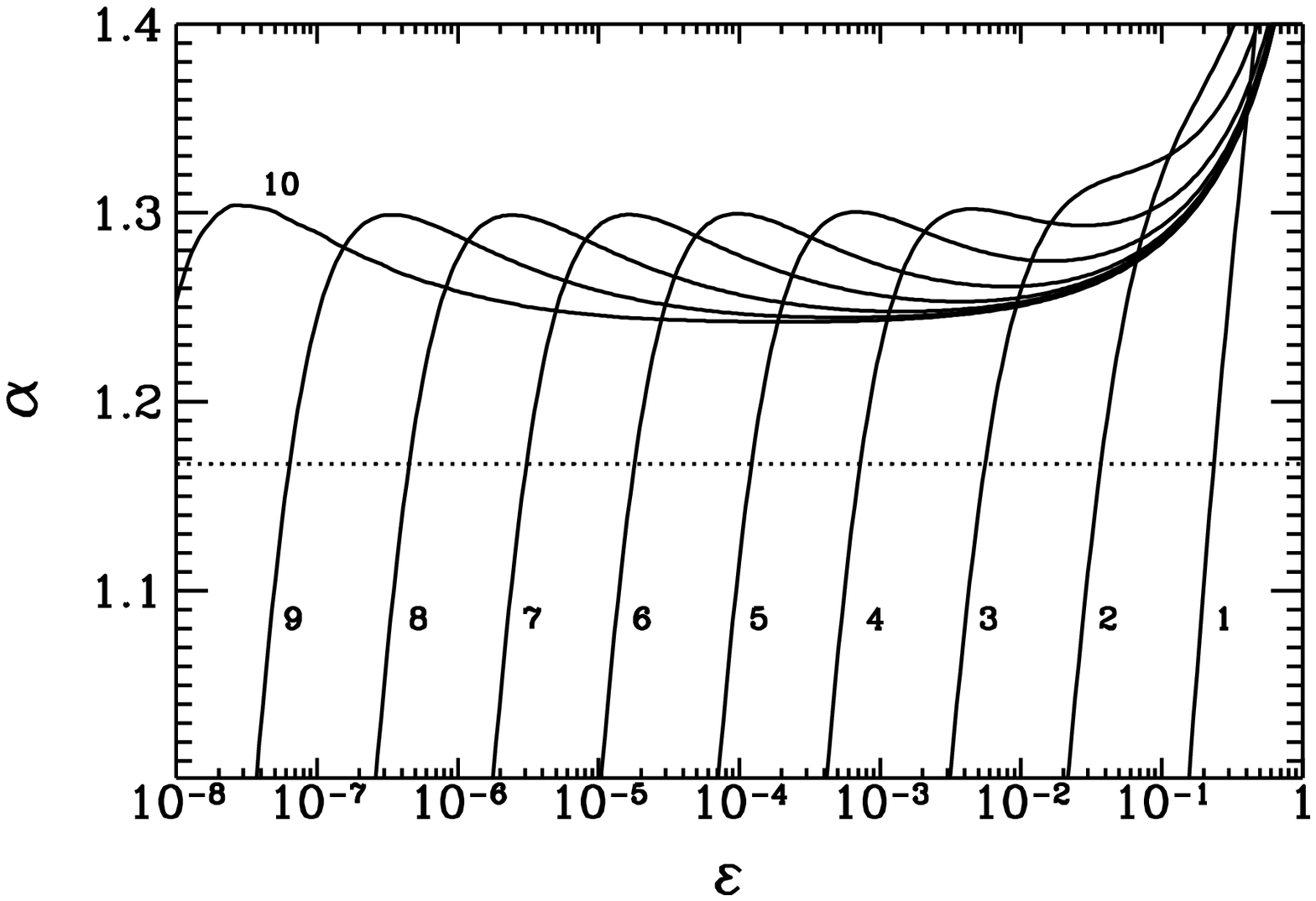,height=4in,width=6in}

\caption{The distribution function (top) and its logarithmic  
derivative
(bottom) are shown at different moments of time prior to condensate  
formation.}
\end{figure}
\begin{figure}

\psfig{file=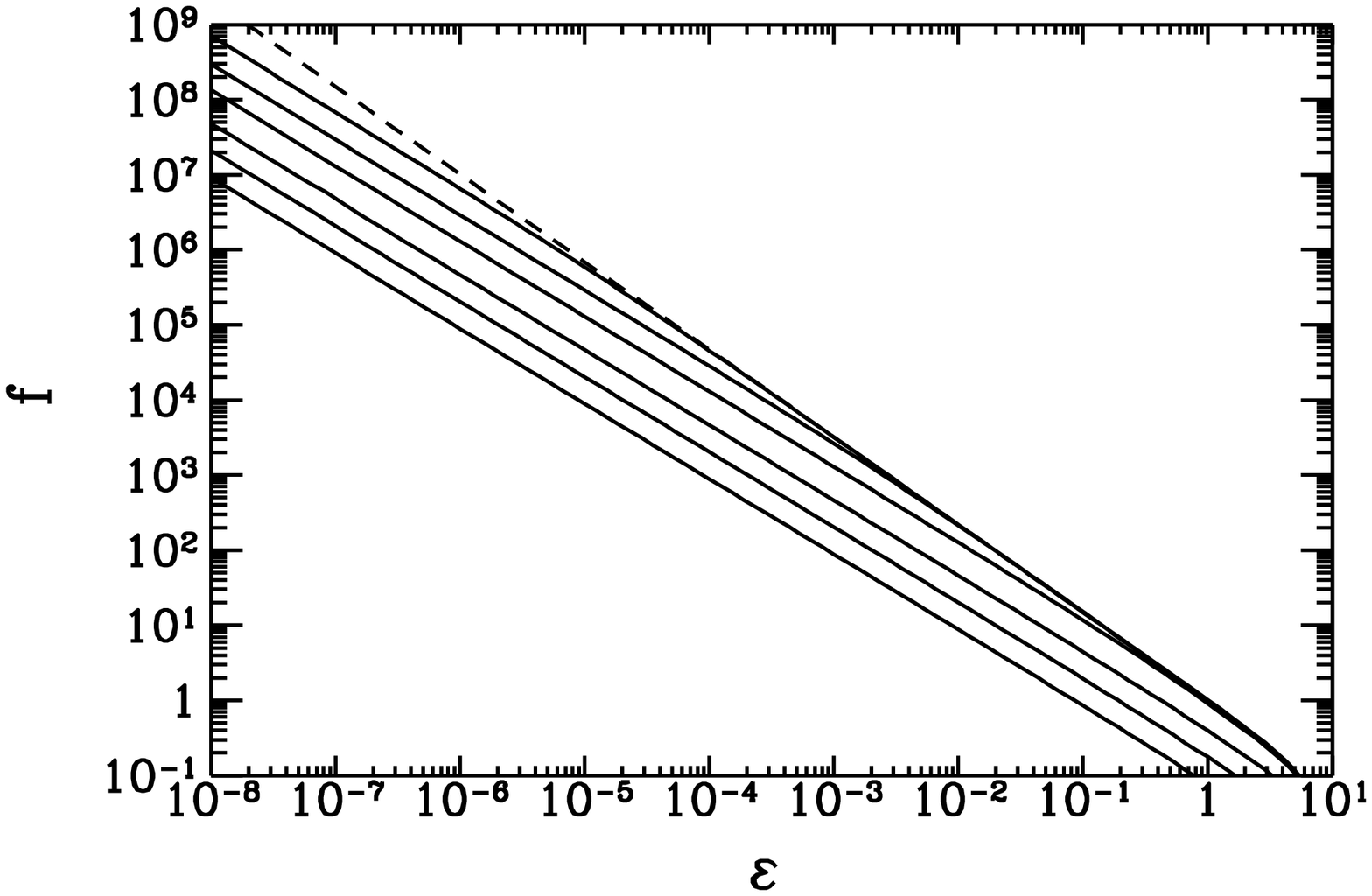,height=4in,width=6in}

\psfig{file=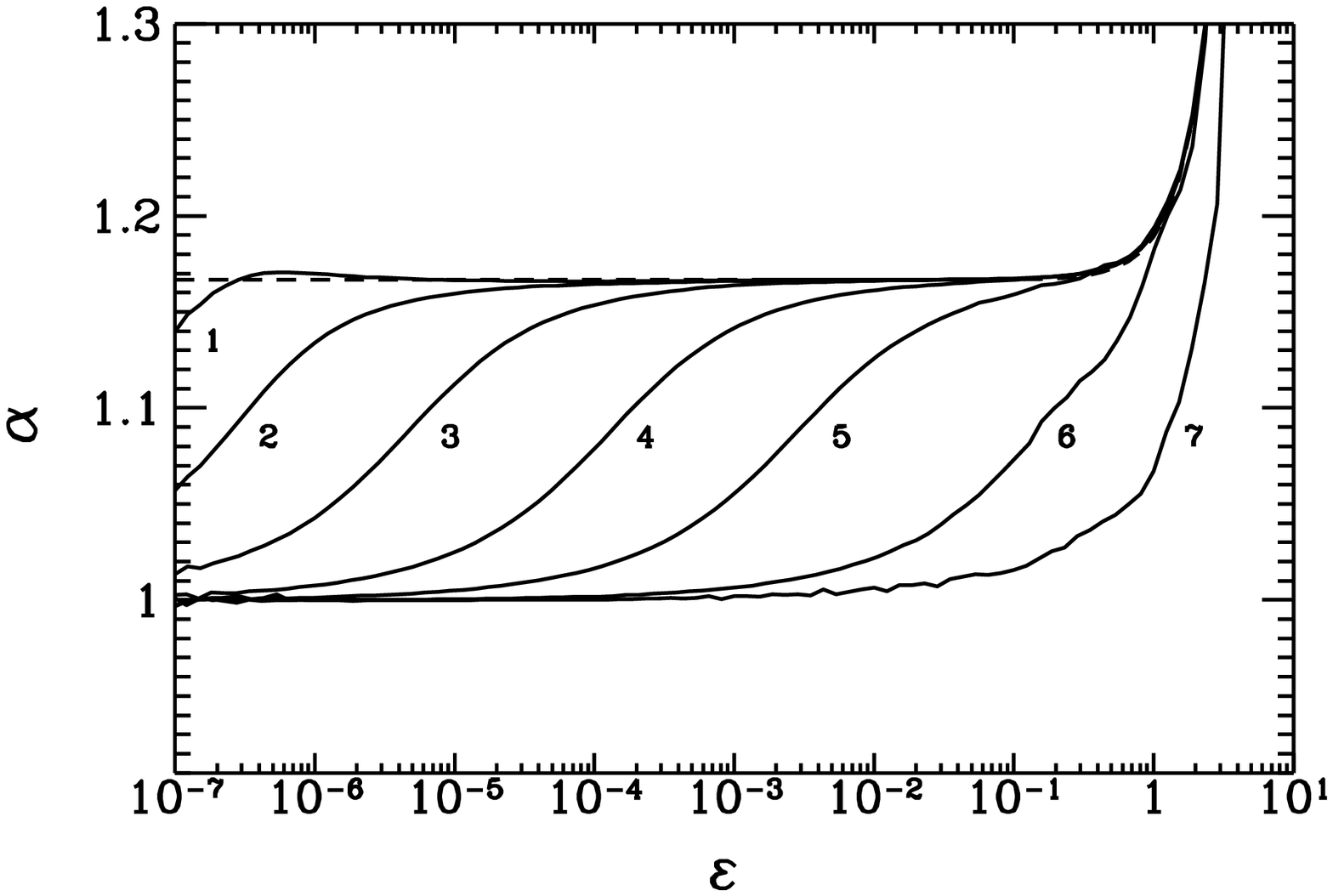,height=4in,width=6in}
\caption{ The distribution function (top) and its logarithmic  
derivative
(bottom) are shown at different moments of time during condensation}
\end{figure}


\begin{references}
\bibitem{bprl}
I. Silveria and J. T. M. Walraven, Prog. Low Temp. Phys.  {\bf 10},  
139 (1986);
H. F. Hess, {\it et. al.,} Phys. Rev. Lett. {\bf 59}, 672 (1987);  R.  
van
Roijen, {\it et. al.,} Phys. Rev. Lett. {\bf 61}, 931 (1988); N.  
Masuhara, {\it
et. al.,} Phys. Rev. Lett. {\bf 61}, 935 (1988).
\bibitem{m90}
C. Monroe, {\it et. al.,} Phys. Rev. Lett. {\bf 65}, 1571 (1990); A.  
J.
Moerdijk and B. J. Verhaar, Phys. Rev. Lett. {\bf 73}, 518 (1994).
\bibitem{it91}
I. I. Tkachev, \pl {\bf B261}, 289 (1991).
\bibitem{kt93}
E. W. Kolb and I. I. Tkachev, Phys. Rev. Lett. {\bf 71}, 3051 (1993);
Phys. Rev. D {\bf 49}, 5040 (1994).
\bibitem{ih76}
M. Inoue and E. Hanamura, J. Phys. Soc. Jpn {\bf 41}, 771 (1976).
\bibitem{ly77}
E. Levich and V. Yakhot, Phys. Rev. B {\bf 15} 243 (1977).
\bibitem{ly78}
E. Levich and V. Yakhot, J. Phys. A {\bf 11}, 2237 (1978).
\bibitem{sw89}
D. W. Snoke and J. P. Wolfe, Phys. Rev. B {\bf 39}, 4030 (1989).
\bibitem{s91}
H. T. C. Stoof, Phys. Rev. Lett. {\bf 66}, 3148 (1991).
\bibitem{kss92}
Yu. M. Kagan, B. V. Svistunov and G. V. Shlyapnikov, Zh. Eksp. Teor.  
Fiz. {\bf
101}, 528 (1992) [Sov. Phys. JETP {\bf 74}, 279 (1992)].
\bibitem{k41}
A. N. Kolmogorov, Dokl. Akad. Nauk SSSR {\bf 30}, 299 (1941).
\bibitem{z66}
V. E. Zakharov, Zh. Eksp. Teor. Fiz. {\bf 51}, 688 (1966)
[Sov. Phys. JETP {\bf 24}, 455 (1966)]; Zh. Eksp. Teor. Fiz. {\bf  
62}, 1745
(1972) [Sov. Phys. JETP {\bf 35}, 908 (1972)].

\end{references}
\end{document}